\documentclass[aps,floatfix,prb,twocolumn,showpacs]{revtex4-1} 
\usepackage{graphicx}

\begin{document}
\title[Exact solution for high harmonic generation \ldots]{Exact solution for high harmonic generation and the response to an ac driving field for a charge density wave insulator}
\author{Wen Shen$^1$, A. F. Kemper$^2$, T. P. Devereaux$^{3,4}$ and J. K. Freericks$^1$}
\address{$^1$Department of Physics, Georgetown University, 37th and O Sts. NW, Washington, DC 20057, USA}
\address{$^2$Computational Research Division, Lawrence Berkeley National Laboratory,
Berkeley, CA 94720, USA}
\address{$^3$Stanford Institute for Materials and Energy Science, SLAC National
Accelerator Laboratory, Menlo Park, CA 94025, USA}
\address{$^4$Geballe Laboratory for Advanced Materials, Stanford University, Stanford,
CA 94305, USA}
\begin{abstract}
We develop and exactly solve a model for electrons driven by  pulsed or continuous ac fields.  The theory includes both the photoexcitation process as well as the subsequent  acceleration of the electrons. In the case of an ac response, we examine both the nonequilibrium density of states and the current. In the case of pulsed light for high harmionic generation, we find the radiated light assumes a nearly universal behavior, with only limited dependence on the parameters of the system, except for the amplitude of the driving field, which determines the range of high harmonics generated and a tendency toward a narrowing of the peaks in a charge density wave versus a metal.  This type of high harmonic generation can potentially be used for the creation of solid-state-based ultrafast light sources.
\end{abstract}
\date{\today}
\maketitle

\section{Introduction}

Recent experimental work has shown that high harmonic generation of light can be seen from a solid material when it is
hit by a large amplitude femtosecond light pulse~\cite{Ghimire}. The light generation comes from the photoexcitation and subsequent acceleration of electrons and holes in the material and the Bloch oscillations that occur as the electrons and holes are accelerated towards the Brillouin zone boundary. Theoretical descriptions of this phenomena have been performed within Boltzmann and semiclassical approaches and within a many-body formulation.  The semiclassical approaches include the dipolar matrix element that couples different symmetry bands together to allow the field to photoexcite electrons and leave behind holes \cite{Ghimire,mucke0,golde1,golde2,mucke}.  The many-body formulation has primarily worked with a single band approach, and hence has not yet examined the photoexcitation process in detail \cite{JimHHG,lex}.  In this work, we examine a simple band model that has a gap due to a checkerboard ordering pattern of the underlying lattice potential and the corresponding charge-density-wave structure that results for the electronic band structure. Because the resulting two bands have the same symmetry and originate from the same single band when the checkerboard potential vanishes, one can directly excite from the lower to the upper band with a formalism that uses just the Peierls substitution to describe the nonlinear effects of the driving electric field (the dipolar matrix element between these bands vanishes since they have the same symmetry). Hence, this model is able to properly describe the whole process for high harmonic generation in the solid within a fully quantum model.  Furthermore, because it is effectively a noninteracting problem, the full nonequilibrium dynamics, including all of these quantum-mechanical effects, can be solved exactly using an efficient parallel algorithm. 

High harmonic generation is an important subject within the optics community because the high
harmonic pulses can be employed to generate sources of light.
These light sources can then be used to selectively drive excitations in specific materials, allowing an experiment to focus on excitation by collective modes with frequencies in a narrow range of energy.
In high harmonics generated from crystals, the field amplitude
of the high harmonics can become comparable to that of the fundamental, making it a possible source
for  light generation \cite{Heblin}.
Furthermore, since the high order harmonics generated from solids are temporally localized
to the center of the driving pulse \cite{lex}, the high order harmonics can have
short durations into the attosecond regime, allowing for ultrashort pulsed light generation.

In this paper, we first describe the formalism needed to solve the nonequilibrium problem exactly in Sec. II. This includes both the formal solution for the Green's functions in the presence of the field, and how to apply the Green's functions to calculate physical observables. In Sec. III, we focus on two different scenarios.  The first is the case of applying an ac driving field.  In this case, one generates a current with the same period as the driving field, but as the amplitude is increased, the current develops significant oscillatory structure within the period, which gives rise to higher harmonic generation.  Performing a Fourier transform of the current, we show that the higher harmonics develop additional oscillatory structure, similar to that of beats.  Next, we consider the case of a pulse driving the system.  Here, we find results that look very similar to those of single-band calculations, indicating that the details for how the photoexcitation takes place does not play a significant role in determining the structure of the high harmonic generation spectra except for an overall narrowing effect on the harmonic peaks.  This is a particularly surprising result, since the simple Landau-Zener theory for tunneling shows that the ability to excite from the lower band to the upper band depends exponentially on the field amplitude, and hence one would expect the photoexcitation to take place primarily at the points where the field amplitude is maximal.  But we find results similar to earlier calculations which did not take the photoexcitation process into account at all \cite{JimHHG,lex}, indicating that the details of the photoexcitation do not play a significant role and instead primarily cause a narrowing effect on the peaks.  This occurs even though the band has Pauli blocking, in that once a state is occupied in the upper band no more electrons can go into that state.  Detailed calculations of the population excited into the upper band as a function of time show that the excitation occurs over an extended region of time, and that for large amplitude pulses, there are also de-excitation processes that reduce the number of electrons in the upper band and repopulate the lower band.  These results are shown elsewhere \cite{amp}. We end the paper by providing our conclusions in Sec. IV.

\section{Formalism}

We study a noninteracting charge-density-wave system on a lattice in the presence of a uniform time-dependent electric field. We work in a gauge that has a spatially uniform, but time dependent vector potential and no scalar potential, so that the system retains translational invariance throughout.  The time-dependent electric field is described via the Peierls' substitution \cite{Peierls}, which is a simplified semi-classical treatment of the electromagnetic field that is exact and nonperturbative. With the Peierls' substitution, the hopping matrix has a time-dependent phase factor \cite{Jauho}
\begin{equation}
t_{ij}(t)= t_{ij}\exp\left[-\frac{ie}{\hbar c}\int_{{\bf R}_i}^{\bf {R}_j}\textbf {A} (\textbf{r},\emph{t})\cdot d\textbf {r}\right].
\end{equation}
From Maxwell's equations, the corresponding electric field $\textbf{E} ({\bf r},\emph{t})$ is found from the derivative of the vector potential $\textbf{A}(\textbf{r}, \emph{t})$.
\begin{equation}
\textbf {E} ({\bf r},\emph{t})=-\frac{1}{c}\frac{\partial {\textbf{A} (\textbf{r},\emph{t})}}{\partial t}.
\end{equation}
The time-dependent Hamiltonian then becomes
\begin{equation}
\mathcal{H}(t)=-\sum_{ i,j}t_{ij}(t)c^\dagger_ic^{}_j+\sum_{i\in A}(U-\mu)c^\dagger_ic^{}_i-\sum_{i\in B}\mu c^{\dagger}_ic^{}_i,
\end{equation}
in the Schroedinger representation. Here, the symbols $A$ and $B$ denote the two sublattices of the bipartite lattice, where we have the checkerboard pattern, and the hopping is only between nearest neighbors. At half filling for the electrons, we have the chemical potential satisfy $\mu=U/2$.  The operators $c^{\dagger}_i$ and $c_i^{}$ denote the creation/destruction operators for a spinless fermion at the $i$th lattice site and they satisfy the canonical fermionic anticommutation relations $\{c^{}_i,c^{\dagger}_j\}_+=\delta_{ij}$.  Note that because these electrons are noninteracting, the spin degree of freedom is trivial to include and hence is neglected here.

In a spatially uniform field directed along the diagonal of the hypercubic lattice in $d$-dimensions, $\textbf {A}(t)=A(t)(1,1,\ldots,1)$, the time-dependent band structure in momentum space for the $U=0$ case is,
\begin{equation}
\varepsilon_k(t)=-\displaystyle\sum_{ij}t_{ij}\exp [-i(\textbf{k}-\frac{e}{\hbar c}\textbf {A} (\emph{t}))\cdot (\textbf{R}_{iA}-\textbf{R}_{jB})].
\end{equation}
So the effect of the Peierls' substitution is to add a time-dependent shift to the momentum in the noninteracting electronic band structure at $U=0$:
\begin{equation}
\varepsilon_k(t)=-\lim_{d\rightarrow\infty}\displaystyle\sum_{l=1}^{d}\frac{t^*}{\sqrt{d}}\cos\left[a\left(k_l-\frac{e{A}(t)}{\hbar c}\right)\right]
\end{equation}
where we have scaled the hopping to give a nontrivial result in the infinite-dimensional limit ($t^*$ will serve as the energy unit and remains finite). Generalizations to other spatial dimensions or to other directions of the electric field are straightforward to do.

Transforming to momentum space, where
\begin{equation}
\emph{c}_{i}^{\dagger}(t)=\displaystyle\sum_{k:\varepsilon_k<0}[e^{-i\textbf{k}\cdot{\textbf{R}_{i}}}\emph{c}_{k}^{\dagger}(t)+e^{-i(\textbf{k}+\textbf{Q})\cdot{\textbf{R}_{i}}}\emph{c}_{k+Q}^{\dagger}(t)]
\end{equation}
with ${\bf Q}=(\pi,\pi,\pi,\ldots )$, the momentum-space Hamiltonian in the Schr$\ddot{\textrm{o}}$dinger representation becomes
\begin{eqnarray}
\mathcal{H}_S(t)&=& \displaystyle\sum_{k:\varepsilon_k<0}\left ( \begin{array}{cc}\emph{c}_{k}^{\dagger}&\emph{c}_{k+Q}^{\dagger}\end{array}\right )\\
&\times&
\left(\begin{array}{cc}{\varepsilon_k(t)+U/2-\mu} & {U/2} \\ {U/2} & {-\varepsilon_k(t)+U/2-\mu}\end{array}\right)
\left(\begin{array}{c}\emph{c}_{k}^{}\\\emph{c}_{k+Q}^{}\end{array}\right)\nonumber
\end{eqnarray}
where the summation is restricted to the small Brillouin zone with $\varepsilon_k<0$ because the reduced translational symmetry (due to the checkerboard ordering of the charge density wave) leads to a smaller Brillouin zone.
The time-dependent band structure $\varepsilon_k(t)$ at $U=0$ can then be expanded with the difference formula of the cosine, 
\begin{equation}
\varepsilon_k(t)= \cos\left(\frac{eaA(t)}{\hbar c}\right)\varepsilon_k+\sin\left(\frac{eaA(t)}{\hbar c}\right)\overline\varepsilon_k
\end{equation}
which depends on the equilibrium band structure at $U=0$
\begin{equation}
\varepsilon_k=-\lim_{d\rightarrow\infty}\displaystyle\sum_{l=1}^d\frac{t^*}{\sqrt{d}}\cos(ak_l)
\end{equation}
and the projection of the equilibrium velocity along the field direction
\begin{equation}
\overline\varepsilon_k=-\lim_{d\rightarrow\infty}\displaystyle\sum_{l=1}^d\frac{t^*}{\sqrt{d}}\sin(ak_l).
\end{equation}

We will be solving for the Green's functions in order to determine the time evolution of the system.  Normally, the Green's functions are defined in terms of the creation and destruction operators in the Heisenberg representation. Because the Hamiltonian is noninteracting, we can directly solve for the Heisenberg operators by solving the equation of motion, which closes at the lowest order.
The operators $c_k(t)$ and $c_{k+Q}(t)$ satisfy
\begin{equation}
i\hbar\frac{dc_k(t)}{dt}=-\left[\mathcal{H}_H(t),c_k(t)\right] 
\end {equation}
and
\begin{equation}
i\hbar\frac{dc_{k+Q}(t)}{dt}=-\left[\mathcal{H}_H(t),c_{k+Q}(t)\right] ,
\end{equation}
where $\mathcal{H}_H(t)$ is the Heisenberg representation for the Hamiltonian.
Substituting in the time-dependent Hamiltonian and evaluating the commutators,
yields
\begin{equation}
i\hbar\frac{dc_k(t)}{dt}=[\varepsilon_k(t)+\frac{U}{2}-\mu]c_k(t)+\frac{U}{2}c_{k+Q}(t)
\end{equation}
\begin{equation}
i\hbar\frac{dc_{k+Q}(t)}{dt}= \frac{U}{2}c_{k}(t)+[-\varepsilon_k(t)+\frac{U}{2}-\mu]c_{k+Q}(t).
\end{equation}
Then the time evolution for the creation and destruction operators satisfies
\begin{equation}
\left(\begin{array}{c}c_{k}(t)\\c_{k+Q}(t)\end{array}\right)=U(k,t,t_0)\left(\begin{array}{c}c_{k}(t_0)\\c_{k+Q}(t_0)\end{array}\right).
\end{equation}
The time-evolution operator $U(k,t,t')$ is a time ordered product for each momentum
\begin{eqnarray}
&~&\emph U(k,t,t')=\\
&~&\emph T_t \exp\left[ -\frac{i}{\hbar}\int_{t'}^t d\overline{t}\left(\begin{array}{cc}{\frac{U}{2}-\mu+\varepsilon_k(\overline{t})} & {\frac{U}{2}} \\ {\frac{U}{2}} &{\frac{U}{2}-\mu-\varepsilon_k(\overline{t})} \end{array}\right)\right].\nonumber
\end{eqnarray}
Since there are time-dependent terms inside the exponential, we must numerically calculate the time evolution operator $U(k,t,t')$.
For a small time step $\Delta t$ at time $t$, we have
\begin{eqnarray}
&~&U(k,t,t-\Delta t)=\\
&~&\exp\left[-\frac{i\Delta t}{\hbar}\left(\begin{array}{cc}{\frac{U}{2}-\mu+\varepsilon_k(t-\frac{\Delta t}{2})} & {\frac{U}{2}} \\ {\frac{U}{2}} &{\frac{U}{2}-\mu-\varepsilon_k(t-\frac{\Delta t}{2})}\end{array}\right)\right],\nonumber
\end{eqnarray}
using a midpoint integration rule.
Since this is the exponential of a $2\times 2$ matrix, it can be determined exactly to be
\begin{eqnarray}
	U(k,t,t-\Delta t)&=&\cos\left (\frac{\Delta t}{\hbar}\sqrt{\varepsilon_k^2\left (t-\frac{\Delta t}{2}\right )+\frac{U^2}{4}}\right )\textbf{I}\nonumber \\ 
&-&i \left(\begin{array}{cc}{\varepsilon_k(t-\frac{\Delta t}{2})} & {\frac{U}{2}} \\ {\frac{U}{2}} &{-\varepsilon_k(t-\frac{\Delta t}{2})}\end{array}\right)\nonumber\\
&\times&
\frac{\sin\left (\frac{\Delta t}{\hbar}{\sqrt{\varepsilon_k^2(t-\frac{\Delta t}{2})+\frac{U^2}{4}}}\right )}{ \sqrt{\varepsilon^2_k(t-\frac{\Delta t}{2})+\frac{U^2}{4}}}.
\end{eqnarray}
Here, we used the result that the chemical potential satisfies $\mu=U/2$ at half-filling to make the equation less cumbersome.
In these calculations, we start from a minimum time $t_0$ producing the time-evolution operator as,
\begin{eqnarray}
U(k,t,t_0)&=&U(k,t,t-\Delta t)U(k,t-\Delta t, t-2\Delta t)\dots \nonumber\\
&\times&U(k,t_0+\Delta t,t_0)
\end{eqnarray}
where the $2\times 2$ matrix structure involves repeated matrix multiplication in this equation.
For each $k$, the two-time evolution operator is found from the identity
\begin{equation}
U(k,t,t')=U(k,t,t_0)U^{\dagger}(k,t_0,t').
\label{eq: evolution_identity}
\end{equation}
Once the time evolution at each time pair is found, we then calculate the nonequilibrium Green's functions to obtain the physical properties of the system. 

The retarded Green's function is defined as follows:
\begin{equation}
G^R_{ij}(t,t')=-i\theta(t-t')\langle \{c^{}_i(t),c^{\dagger}_j(t')\}_+\rangle,
\end{equation}
with $\langle \hat\mathcal{O} \rangle={\rm Tr}\exp[-\beta\mathcal{H}(t\rightarrow -\infty)]\hat\mathcal{O} /\mathcal{Z}$ (for any operator $\hat\mathcal{O}$ and the equilibrium partition function $\mathcal{Z}={\rm Tr}\exp[-\beta\mathcal{H}(t\rightarrow -\infty)]$. Here $\beta=1/T$ is the inverse temperature that the system is initialized in before the field is turned on (in this work we will always start the system at $T=0$ or $\beta=\infty$). Substituting in the time-evolution operators for the momentum-dependent operators produces for the local Green's function
\begin{widetext}
\begin{eqnarray}
G^R_{ii} (t,t')&=& -i \theta (t-t')\sum\limits_{k:\varepsilon_k<0}\langle \{c_k(t_0)U_{11}(k,t,t_0)+c_{k+Q}(t_0)U_{12}(k,t,t_0)\pm c_k(t_0)U_{21}(k,t,t_0)\pm c_{k+Q}(t_0)U_{22}(k,t,t_0),\nonumber\\
&~&c^{\dagger}_k(t_0)U^{\dagger}_{11}(k,t_0,t')+c^{\dagger}_{k+Q}(t_0)U^{\dagger}_{21}(k,t_0,t')\pm c^{\dagger}_k(t_0)U^{\dagger}_{12}(k,t_0,t')\pm c^{\dagger}_{k+Q}(t_0)U^{\dagger}_{22}(k,t_0,t')\}_+\rangle .
\end{eqnarray}
We assume $t_0$ represents a time far before the time that the external field is turned on, so that the system is in equilibrium at time $t_0$. The plus sign is for the $A$ sublattice and the minus sign is for the $B$ sublattice
because there is a phase shift $e^{\pm i\vec{\bf Q}\cdot{\bf R}_A}=1$ and $e^{\pm i\vec{\bf Q}\cdot {\bf R}_B}=-1$ . 
Here $U_{ab}(k,t,t')$ and $U^{\dagger}_{ab}(k,t,t')$ represent the elements of row $a$ and column $b$ in the $U(k,t,t')$ and $U^{\dagger}(k,t,t')$ matrices, respectively. 
With the anticommutation relations for the fermionic operators, the retarded Green's function simplifies to
\begin{eqnarray}
G^R_{ii} (t,t')&=&-i \theta (t-t')\sum_{k: \varepsilon_k<0}[U_{11}(k,t,t_0)\pm U_{21}(k,t,t_0)][U^{\dagger}_{11}(k,t_0,t')\pm U^{\dagger}_{12}(k,t_0,t')]
+[U_{12}(k,t,t_0)\pm U_{22}(k,t,t_0)]\nonumber\\
&\times&[U_{21}^{\dagger}(k,t_0,t')\pm U^{\dagger}_{22}(k,t_0,t')].
\end{eqnarray}
\end{widetext}
Next, we find that the local retarded Green's function becomes just a function of the elements of the time-evolution operator between the two times $t$ and $t'$:
\begin{eqnarray}
G^R_{ii} (t,t')&=&-i \theta (t-t')\displaystyle\sum_{k: \varepsilon_k<0}[U_{11}(k,t,t')+U_{22}(k,t,t')\nonumber\\
&\pm& U_{12}(k,t,t')\pm U_{21}(k,t,t')],
\end{eqnarray}
which follows from the fundamental identity of the evolution operators in Eq.~(\ref{eq: evolution_identity}).
From the above equation, we can see that the retarded Green's function only depends on the evolution between the two times $t$ and $t'$ and not on the previous history of the evolution of the system. This is because the retarded Green's function determines the character of the quantum states of the system, which is determined by the current Hamiltonian, and not on the complete history of the evolution of the system.

Now we employ the average and relative time coordinates which were introduced by Wigner for nonequilibrium Green's functions \cite{Wigner}. The relative and average times are defined via,
\begin{equation}
t_{rel}=t-t', \ \ t_{ave}=\frac{t+t'}{2},
\end{equation}
respectively. The local retarded Green's function in the frequency domain is the Fourier transform of the local retarded Green's function with respect to the relative time for a fixed average time:
\begin{equation}
G^R_{ii} (\omega,t_{ave})=\int dt_{rel} e^{i\omega t_{rel}}G^R_{ii}(t_{rel},t_{ave}).
\end{equation}
From this Green's function, one can construct the nonequilibrium local density of states
\begin{equation}
A_{i}(\omega,t_{ave})=-\frac{1}{\pi}{\rm Im}G^R_{ii}(\omega,t_{ave}).
\end{equation}

In order to calculate the current as a function of time, we need to use the lesser Green's function, which determines how the quantum states are filled.
The lesser Green's function is defined as
\begin{equation}
G^<_{ij}(t,t')=i\langle c^{\dagger}_j(t') c^{}_i(t)\rangle,
\end{equation}
with the operators in the Heisenberg representation. 
Unlike the retarded Green's function, the lesser Green's function does not simplify to depend only on the relative time evolution from $t$ to $t'$, but it also depends on the history of the time evolution. We need the lesser Green's functions in momentum space, which has the $2\times 2$ matrix structure discussed above. 
Here we write down these four components of the lesser Green's functions:
\begin{widetext}
\begin{eqnarray}
G^{<}_{11}(k,t,t')&=&i\langle c_k^{\dagger}(t')c_k(t)\rangle\\ &=& U_{11}^{\dagger}(k,t_0,t')U_{11}(k,t,t_0)\langle c_k^{\dagger}(t_0)c_k(t_0)\rangle +U_{11}^{\dagger}(k,t_0,t')U_{12}(k,t,t_0)\langle c_k^{\dagger}(t_0)c_{k+Q}(t_0)\rangle \nonumber\\ &+&U_{21}^{\dagger}(k,t_0,t')U_{11}(k,t,t_0)\langle c_{k+Q}^{\dagger}(t_0) c_k(t_0)\rangle+U_{21}^{\dagger}(k,t_0,t')U_{12}(k,t,t_0) \langle c_{k+Q}^{\dagger}(t_0)c_{k+Q}(t_0) \rangle \nonumber\\
G^{<}_{12}(k,t,t')&=&i\langle c_{k+Q}^{\dagger}(t')c_k(t)\rangle\\ &=& U_{11}^{\dagger}(k,t_0,t')U_{21}(k,t,t_0)\langle c_k^{\dagger}(t_0)c_k(t_0)\rangle +U_{11}^{\dagger}(k,t_0,t')U_{22}(k,t,t_0)\langle c_k^{\dagger}(t_0)c_{k+Q}(t_0)\rangle \nonumber\\ &+&U_{21}^{\dagger}(k,t_0,t')U_{21}(k,t,t_0)\langle c_{k+Q}^{\dagger}(t_0) c_k(t_0)\rangle+U_{21}^{\dagger}(k,t_0,t')U_{22}(k,t,t_0) \langle c_{k+Q}^{\dagger}(t_0)c_{k+Q}(t_0) \rangle \nonumber\\
G^{<}_{21}(k,t,t')&=&i\langle c_k^{\dagger}(t')c_{k+Q}(t)\rangle\\ &=& U_{12}^{\dagger}(k,t_0,t')U_{11}(k,t,t_0)\langle c_k^{\dagger}(t_0)c_k(t_0)\rangle +U_{22}^{\dagger}(k,t_0,t')U_{12}(k,t,t_0)\langle c_k^{\dagger}(t_0)c_{k+Q}(t_0)\rangle \nonumber\\ &+&U_{12}^{\dagger}(k,t_0,t')U_{11}(k,t,t_0)\langle c_{k+Q}^{\dagger}(t_0) c_k(t_0)\rangle+U_{22}^{\dagger}(k,t_0,t')U_{12}(k,t,t_0) \langle c_{k+Q}^{\dagger}(t_0)c_{k+Q}(t_0) \rangle \nonumber\\
G^{<}_{22}(k,t,t')&=&i\langle c_{k+Q}^{\dagger}(t')c_{k+Q}(t)\rangle\\ &=& U_{12}^{\dagger}(k,t_0,t')U_{21}(k,t,t_0)\langle c_k^{\dagger}(t_0)c_k(t_0)\rangle +U_{22}^{\dagger}(k,t_0,t')U_{22}(k,t,t_0)\langle c_k^{\dagger}(t_0)c_{k+Q}(t_0)\rangle \nonumber\\ &+&U_{12}^{\dagger}(k,t_0,t')U_{21}(k,t,t_0)\langle c_{k+Q}^{\dagger}(t_0) c_k(t_0)\rangle+U_{22}^{\dagger}(k,t_0,t')U_{22}(k,t,t_0) \langle c_{k+Q}^{\dagger}(t_0)c_{k+Q}(t_0) \rangle . \nonumber
\end{eqnarray}
\end{widetext}
As before, the subscript 1 denotes the {\bf k} element and the subscript 2 denotes the ${\bf k}+{\bf Q}$ element of the $2\times 2$ momentum-dependent Green's function matrix.

We take our starting time $t_0$ as a time well before the time we turn on the electric field. We assume the system is in equilibrium at $t_0$ and start the evolution from there. To determine the expectation values of the different momentum operators, we must convert those expectation values into the equilibrium distributions found from diagonalizing the equilibrium Hamiltonian into its two respective bands.
The equilibrium Hamiltonian is first written in the $2\times2$ $(c_k, c_{k+Q})$ basis,
\begin{eqnarray}
&~&\mathcal{H}(t\rightarrow -\infty)= \displaystyle\sum_{k:\varepsilon_k<0}\left(\begin{array}{cc}\emph{c}_{k}^{\dagger}&\emph{c}_{k+Q}^{\dagger}\end{array}\right)\\
&\times&
\left(\begin{array}{cc}{\varepsilon_k+U/2-\mu} & {U/2} \\ {U/2} & {-\varepsilon_k+U/2-\mu}\end{array}\right)
\left(\begin{array}{c}\emph{c}_{k}\\\emph{c}_{k+Q}\end{array}\right).\nonumber
\end{eqnarray}
The Hamiltonian is then diagonalized with the following eigenfunction basis,
\begin{equation}
c_{k+}=\alpha_k\emph{c}_{k}+\beta_k\emph{c}_{k+Q}
\end{equation}
\begin{equation}
c_{k-}=\beta_k\emph{c}_{k}-\alpha_k\emph{c}_{k+Q}. 
\end{equation}
With $\alpha_k$ and $\beta_k$ numbers, we can find that $\emph{c}_{k+}$ and $\emph{c}_{k-}$ are annihilation operators for the upper band and lower band eigenstates at the momentum point $k$ in the reduced Brillouin zone. 
Here $\alpha_k$ and $\beta_k$ satisfy
\begin{equation}
\alpha_k=\frac{\frac{U}{2}}{\sqrt{2\left (\varepsilon_k^2+\frac{U^2}{4}-\varepsilon_k\sqrt{\varepsilon_k^2+\frac{U^2}{4}}\right )}}
\end{equation}
and
\begin{equation}
\beta_k=\frac{-\varepsilon_k+\sqrt{\varepsilon_k^2+\frac{U^2}{4}}}{\sqrt{2\left (\varepsilon_k^2+\frac{U^2}{4}-\varepsilon_k\sqrt{\varepsilon_k^2+\frac{U^2}{4}}\right )}}.
\end{equation}
We can also rewrite the Hamiltonian in the $2\times 2$ $ (\emph{c}_{k+},\emph{c}_{k-})$ basis as
\begin{eqnarray}
&~&\mathcal{H}(t\rightarrow -\infty)=\sum_{k:\varepsilon_k<0}\left(\begin{array}{cc}\emph{c}_{k+}^{\dagger}&\emph{c}_{k-}^{\dagger}\end{array}\right)\nonumber\\
&\times&\left(\begin{array}{cc}{\varepsilon_{k+}-\mu} & {0} \\ {0} & {\varepsilon_{k-}-\mu}\end{array}\right) \left(\begin{array}{c}\emph{c}_{k+}\\\emph{c}_{k-}\end{array}\right).
\end{eqnarray}
Here, $\varepsilon_{k+}$ and $\varepsilon_{k-} $ are the two charge-density-wave bands
\begin{equation}
\varepsilon_{k\pm}=\frac{U}{2} \pm \sqrt{\varepsilon_k^2+\frac{U^2}{4}}.
\end{equation}
From the energy spectrum, we can determine the equilibrium density of states. For example, in the infinite dimensional case the average density of states is
\begin{equation}
A(\omega)=\rho \left (-\frac{1}{t^*}\sqrt{\omega^2-\frac{U^2}{4}}\right ) \frac{|\omega|}{\sqrt{\omega^2-\frac{U^2}{4}}},
\end{equation}
with $\rho (\varepsilon)=\exp(-\varepsilon^2)/\sqrt{\pi}$.

We take the state at $t_0$ to be the zero-temperature equilibrium state
\begin{eqnarray}
\langle c_k^{\dagger}(t_0)c_k(t_0)\rangle &=&\alpha_k^2 \langle c_{k+}^{\dagger}(t_0)c_{k+}(t_0)\rangle \nonumber\\
&+&\beta_k^2 \langle c_{k-}^{\dagger}(t_0)c_{k-}(t_0)
=\beta_k^2 \\
\langle c_{k+Q}^{\dagger}(t_0)c_{k+Q}(t_0)\rangle &=&\alpha_k^2\langle c_{k-}^{\dagger}(t_0)c_{k-}(t_0)\rangle \nonumber\\
&+&\beta_k^2 \langle c_{k+}^{\dagger}(t_0)c_{k+}(t_0)\rangle 
=\alpha_k^2\\
\langle c_k^{\dagger}(t_0)c_{k+Q}(t_0)\rangle &=&\alpha_k\beta_k (\langle c_{k+}^{\dagger}(t_0)c_{k+}(t_0)\rangle \nonumber\\
&-&\langle c_{k-}^{\dagger}(t_0)c_{k-}(t_0)\rangle ) 
=-\alpha_k\beta_k \\
\langle c_{k+Q}^{\dagger}(t_0)c_k(t_0)\rangle &=&\alpha_k\beta_k(\langle c_{k+}^{\dagger}(t_0)c_{k+}(t_0)\rangle \nonumber\\
&-&\langle c_{k-}^{\dagger}(t_0)c_{k-}(t_0)\rangle )
=-\alpha_k\beta_k.
\end{eqnarray}
These results follow from the initial occupancies
\begin{equation}
\langle c_{k+}^{\dagger}(t_0)c_{k+}(t_0)\rangle =\frac{1}{1+e^{ (\varepsilon_{k+}-\mu)/T}}=0
\end{equation}
and 
\begin{equation}
\langle c_{k-}^{\dagger}(t_0)c_{k-}(t_0)\rangle =\frac{1}{1+e^{(\varepsilon_{k-}-\mu)/T}}=1
\end{equation} 
since $\mu=U/2$ and $T=0$ (and we have set the Boltzmann constant to 1).

The formula for the current is tedious, but straightforward to derive. It follows just as the derivation does in equilibrium, but one needs to take into account the Peierls' substitution. After some long algebra, one finds that each spatial component of the current becomes
\begin{equation}
\langle j_{\alpha}(t)\rangle=\displaystyle\sum\limits_{k:\varepsilon(k)<0}\frac{1}{\hbar}\nabla_{k_\alpha}\varepsilon(k-eA(t))[ {G}^<_{11}(k,t,t)-{G}^<_{22}(k,t,t)].
\end{equation}

Note that we are solving this problem in the infinite-dimensional limit, but the solution is an exact solution in any dimension, with the corresponding changes to a number of formulas to take into account the finite dimensionality of the system.

\section{Numerical implementation}
We solve this model in the infinite-dimensional case at half filling with the units taken as $e=\hbar=a=c=t^*=1$. 
In the limit of the infinite dimensions,  the joint density of states for the $U=0$ case satisfies \cite{bloch_nonint}
\begin{equation}
\rho(\varepsilon_k , \overline\varepsilon_k)=\frac{1}{\pi }\exp\left[-(\varepsilon_k ^2+\overline\varepsilon_k ^2)\right].
\end{equation}
Here the band structure $\varepsilon_k$ and projection of the band velocity on the direction of the field $\overline\varepsilon_k$ are functions of $k$. Special attention needs to be paid to the points when $\varepsilon_k=0$. We only take $1/2$ of the weight because those momentum points are on the boundary of the reduced Brillouin zone and each equivalent momentum point appears twice.

The accuracy of this calculation is both dependent on the time step $\Delta t$ and energy step $\Delta\varepsilon_k$ ($\Delta\overline\varepsilon_k$).
The error caused by $\Delta t$ can be evaluated with the accuracy of the moments of the retarded Green's function. We find this error is less than $10^{-4}$ when we take $\Delta t<0.02$. Here we use $\Delta t=0.01$.
The error caused by the energy step influences the maximum relative time to which we can calculate.
With $\Delta\varepsilon$ =0.005, we can achieve an accurate lesser Green's function as far as $|t_{rel}|<1300$.

\section{AC response}

\begin{figure}[htb]
\centerline{\includegraphics[width=3.5in]{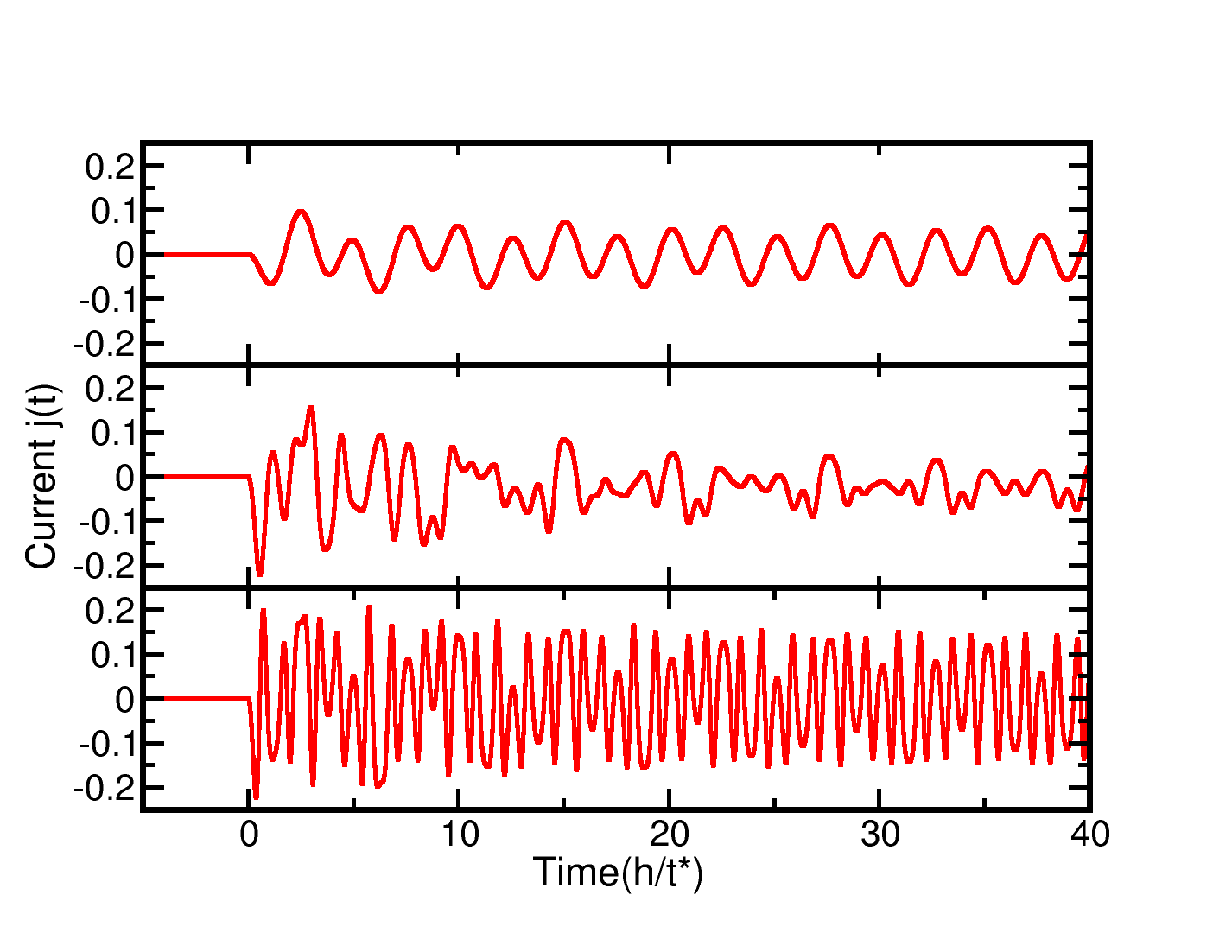}}
\caption{(Color online.) Driven ac current with $U=1$, $\omega_0$=2.5 and different monochromatic ac field amplitudes: (i) $E_0=1$ (top); (ii) $E_0=5$ (center); and (iii) $E_0=10$ (bottom). } \label{fig: ac_current}
\end{figure}

We begin by considering the response of the charge-density wave system to a monochromatic ac field of amplitude $E_0$ and frequency $\omega_0$ applied at $t=0$ [$\mathcal{E}(t)=\theta(t)E_0\sin(\omega_0 t)$].
There are three important energies that can determine frequency-dependent responses of the system: (i) the driving frequency $\omega_0$; (ii) the field amplitude $E_0$, and (iii) the gap of the charge-density-wave system $U$.
Figure~\ref{fig: ac_current}, gives an example of the current response to an ac field. Note how the current starts with a transient response which fairly rapidly approaches a periodic ``steady state'' behavior.  The current is modified from its Bloch oscillating form due to dephasing effects in the driven system, since there is no damping. In this figure, the field amplitude plays an important role in determining the oscillatory response once the field amplitude becomes large enough. It does this by creating additional structure inside the period of the oscillating motion determined by the driving frequency. This more complicated structure can be seen in a Fourier transform of the data as giving rise to higher harmonics in the power spectrum (see below).

\begin{figure} 
\centerline{\includegraphics[width=3.5in]{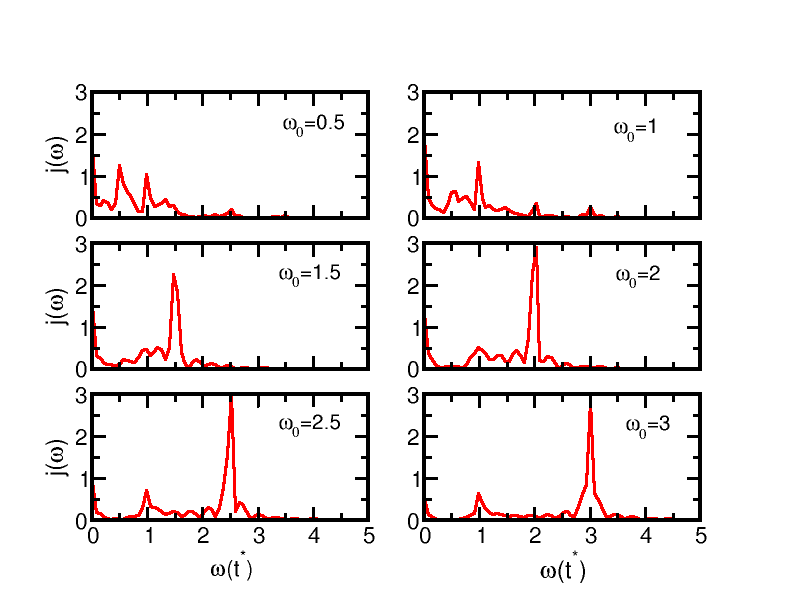}}
\caption{(Color online.) Fourier transform of the ac current for $U=1$ and $E_0$=1 and different $\omega_0$. Note how most panels have peaks at $\omega=1$ and $\omega=\omega_0$. When $\omega_0=1$ higher harmonics are also seen.} \label{fig: ac_current_w}
\end{figure}

Of course, the electrons must follow the ac field and oscillate at a frequency $\omega_0$. In addition, we expect the system to also oscillate at the gap $U$ and at the driving field amplitude $E_0$. We examine the case with $U=1$ and $E_0=1$ for different ac driving frequencies.  Figure~\ref{fig: ac_current_w} shows the Fourier transform of the current over the time interval $[0,80]$ for this case. We are forced into considering a finite time interval, because the simulation is limited by how far in time it can be evolved.  Because we are using a finite time window, peaks that might be sharp over a larger window will become broadened due to the truncation in the time domain.  Nevertheless, this method is quite good for determining what frequencies are providing the dominant contributions to the time trace. We see that peaks are located at $U=E_0$ and $\omega_0$. When $\omega_0$ is away from $U=E_0$, the signal at $U=E_0$ is more significant. It is interesting to note that in the case $\omega_0=U=E_0=1$, peaks are observed at harmonics of the fundamental frequency.

\begin{figure} 
\centerline{\includegraphics[width=3.5in]{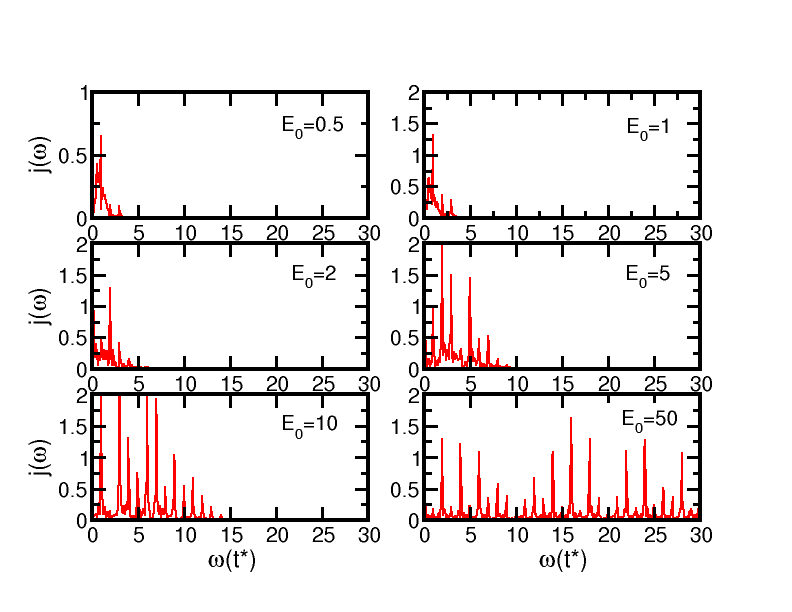}}
\caption{(Color online.) Fourier transform of the ac current for $U=1$ and $\omega_0$=1 with different $E_0$. } \label{fig: ac_current_e0}
\end{figure}

Next, we examine the case that fixes the charge density wave gap $U=1$ and field driving frequency $\omega_0=1$ and varies the field amplitude. This result is plotted in Figure~\ref{fig: ac_current_e0}.
Note how increasing the field amplitude increases the range of harmonic frequencies that are generated in the Fourier transform of the current. This is a direct result of Bloch oscillations, which create high harmonics as the electrons are Bragg reflected at the small Brillouin zone boundary. 
More interesting, is the fact that the power spectrum is not monotonic, but has a beat-like structure to the amplitude of the different harmonic signals. This structure appears to get more complex as the field amplitude is increased.

\begin{figure} 
\centerline{\includegraphics[width=3.5in]{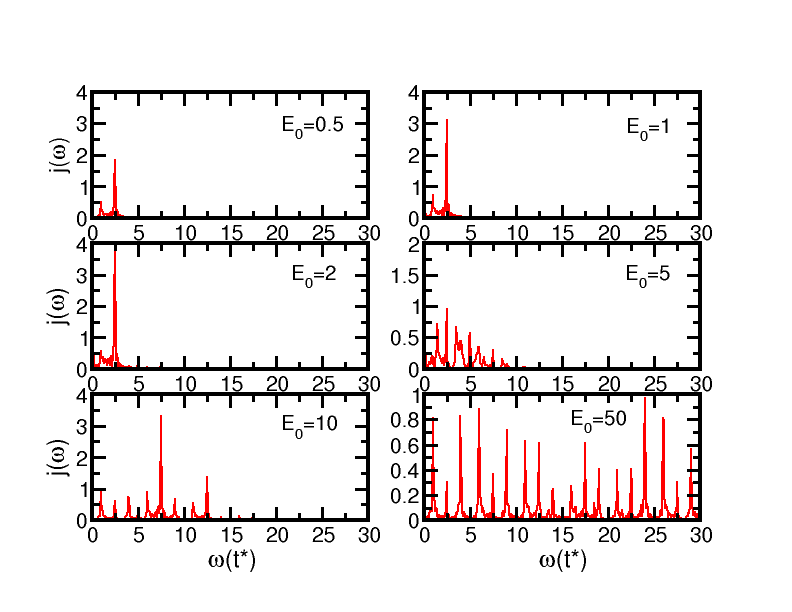}}
\caption{(Color online.) Fourier transform of the ac current for $U=1$, $\omega_0$=2.5 with different $E_0$. Higher harmonics of both frequencies are seen, and joint harmonics are sometimes enhanced, sometimes reduced. } \label{fig: ac_current_e02}
\end{figure}

We also calculate the case with fixed $U=1$ and $\omega_0=2.5$ and varying field amplitude $E_0$. The result for this case is shown in Figure~\ref{fig: ac_current_e02}. 
In this case, we do not see significant harmonics when the field amplitude is low ($E_0$=0.5 ,1, and 2). We see that from $E_0=0.5 $ to $2$, there are only peaks at $U$ and $\omega_0$. Higher harmonics begin to emerge when $E_0=5$. As the amplitude is further increased, it becomes clear that we see harmonics at both frequencies $U$ and $\omega_0$. But, in cases when a frequency is a high harmonic of both $U$ and $\omega_0$, we find that there can be either constructive or destructive interference of the higher harmonic peaks.

\begin{figure} 
\centerline{\includegraphics[width=3.5in]{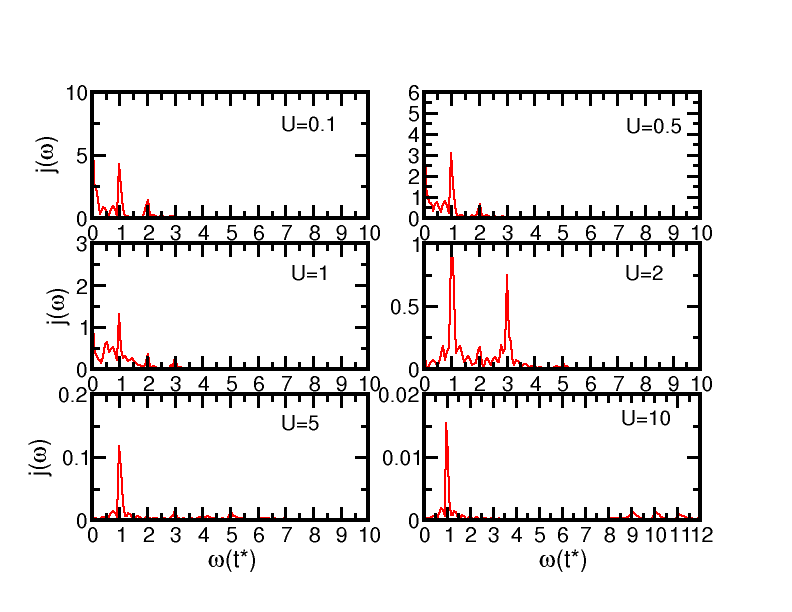}}
\caption{(Color online.) Fourier transform of the ac current for $\omega_0$=1 and $E_0=1$ with different $U$. Note the change in the vertical scale for lower plot. } \label{fig: ac_current_u}
\end{figure}

Finally, we study the case with fixed $E_0=1$ and $\omega_0=1$, but changing $U$ in Figure \ref{fig: ac_current_u}.
When $U$ increases, we see the current amplitude decreases because of the growing insulator gap magnitude. Peaks are observed at frequencies $\omega=U$ and $\omega=E_0=\omega_0$. The field amplitude is too small to generate significant high harmonics. The largest harmonics are generated when $U$ is near 1.

In addition to examining the current traces as functions of time, or their Fourier transform as functions of frequency, we can also examine the ``steady state'' density of states of the system that sets in for long average times.  In equilibrium, the density of states has square-root singularities at the upper and lower band edges.  These singularities give rise to long tails in the time domain, which decay like the inverse square root of the time. When we examine the retarded Green's function for long average times, the two times in the Green's function either correspond to both times being after the field was turned on, or one time before and one after.  When both times are after the field has turned on, the system is determined by its behavior as a quantum system in the presence of the driving field, which will determine the ``steady state'' density of states. In the mixed case of one time before and one after, the system interpolates the Green's function between the equilibrium and the nonequilibrium limits.  In general, however, the tails in the time domain are shorter than in equilibrium, because the nonequilibrium system does not have power law singularities in the density of states. Because the ac field oscillates in time, we expect the density of states to have a dependence on the average time $t_{ave}$.  Averaging the results for the density of states over this period is employed to construct the time-averaged nonequilibrium density of states.

\begin{figure}
\centerline{\includegraphics[width=3.9in]{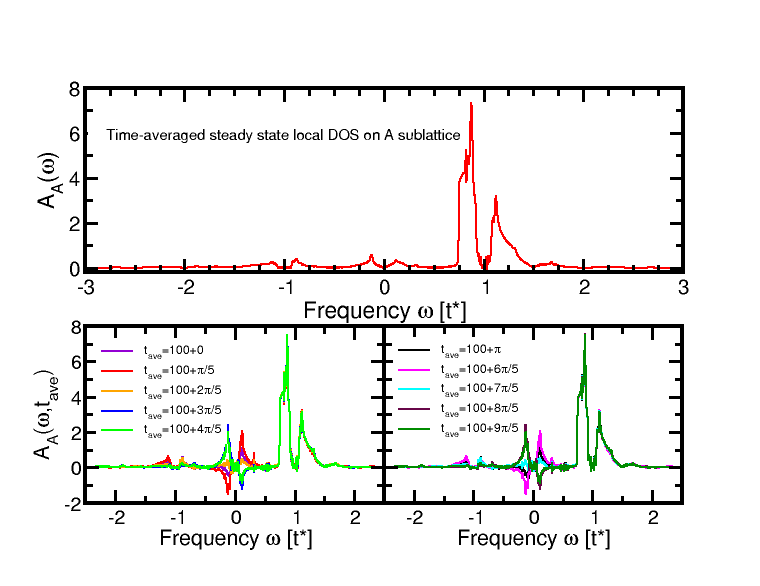}}
\caption{(Color online.) Long-time local density of states for the $A$ sublattice and the case where the field is given by $\mathcal{E}(t)=\theta(t)\sin (t)$ and $U=1.5$. The upper panel shows the time-averaged result, while the lower two panels show the results for different times during the period. We examine the period of $2\pi$ starting from the average time of $t_{ave}=100$.\label{fig: ac_dos}}
\end{figure}

This is done in Figure~\ref{fig: ac_dos}. The top panel shows the time-averaged local density of states for the $A$ sublattice. In a noninteracting metal, we expect to see features at harmonics of the driving frequency, which will occur at the integers for $\omega_0=1$.  Indeed, we do tend to see a sharp reduction of the density of states there, and a smoothing out of all singularities.  The most prominent feature is near $\omega=1$, with other small integers showing smaller features.  This region is where the equilibrium density of states showed the divergence at the inner band edge on the $A$ sublattice.  Near the other small integers, additional features can be seen, but they tend to be small.  Looking at the instantaneous density of states at the different time steps shows that $\omega=0$ and $\omega=-1$ have significant structures for specific times during the period, but they are greatly reduced when averaged over time. In fact, the averaging of the density of states produces the cancellation already over one half period, as one can see that the right lower panel is the same as the left lower panel. The results for the $B$ sublattice are just the mirror image of the top panel in Figure~\ref{fig: ac_dos}. It is likely that the time-averaged DOS should relate directly to the DOS that one can calculate within a Floquet formalism.

\section{High harmonic generation}

Next, we examine the phenomenon of high harmonic generation in solids. In an experimental set-up, this will correspond to applying a large amplitude pulsed field rather than a continuous-beam ac field to the sample. The resulting Bloch oscillations have higher harmonics due to the nonlinear nature of the system when the driving field amplitude is large.

Harmonic generation of light was observed in crystalline quartz by Franken in 1961 after the invention of laser \cite{laser}. In this experiment, a monochromatic laser with a wavelength of 694 nm was focused onto crystalline quartz, and ultraviolet light with a wavelength of 347 nm, or twice the frequency of the original laser, was created. This result can be simply interpreted as the doubling of the frequency of light as two photons combine to one in a nonlinear process.

High harmonic generation was next observed in a plasma generated from a solid material in 1977 \cite{HHG1}, and later high harmonic generation was also observed in a gas. High order harmonic generation from the gas phase has complicated features and is still not fully understood. One of the prevailing theories for high harmonic generation in gases is the three-step model. This model considers the nonlinear optical process as (i) tunnel ionization of an electron, (ii) electron acceleration in the laser field away from the ion and then back toward the ion, and (iii) recombination to its parent ion with an energy release in the form of higher energy photons \cite{Corkum,HHG2,HHG3}.

Recently Ghimire and collaborators reported the observation of high-order harmonic generation in a ZnO crystal, which occurs in the solid due to Bloch oscillations of photoexcited electrons \cite{Ghimire}. In their experiments, they focus linear polarized high-power,  9 cycle mid-infrared laser pulses (3.2-3.7 $\mu$m) onto a single-crystal ZnO wafer near normal incidence. They observed high harmonic peaks beyond the band edge. This observation is fundamentally different from the gas-phase high harmonic generation. Instead of the ionization and recombination of an electron, this phenomenon comes from the acceleration of the electrons in the solid and electron-hole pair excitation and de-excitation, essentially ocurring due to Bloch oscillations.
Though the high harmonic generation is hard to observe in the lattice, it is very natural to obtain the high harmonic generation in theory in the solid from these Bloch oscillations. When a time-dependent electric field $\mathcal{E}(t)$ is applied to the system, the electrons accelerate in momentum space, and when electrons reach the boundary of the Brillouin zone, Bragg reflections occur to fold them back into the first Brillouin zone. This acceleration process  make electrons perform periodic motion in momentum space. If the electric field is strong enough to have $|ae\mathcal{E}|$ large compared to the width of the electronic band, and the driving frequency is small, we would observe a discrete energy spectrum since the electron wave functions become localized. This kind of discrete energy spectrum is called the Wannier-Stark ladder \cite{wannier}. However, when the driving frequency is large, the oscillating charged electrons will radiate light, and the radiation pattern is determined by harmonics of the driving frequency.

\begin{figure}[!ht]
\begin{center}
\includegraphics[width=3.5in]{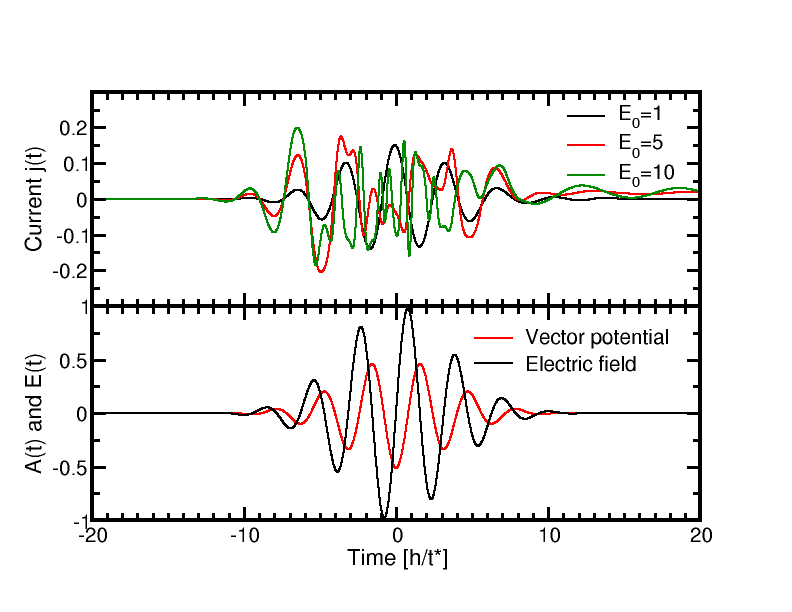}
\end{center}
\caption{(Color online.) Electrical current (top) in the charge-density-wave insulator with $U=1$ and generated by the electric field pulse $\mathcal{E}(t)={E}_0\sin(2t)\exp (-t^2/25)$ (bottom).}
\label{fig: hhg_sin}
\end{figure}
\begin{figure}[!ht]
\begin{center}
\includegraphics[width=3.5in]{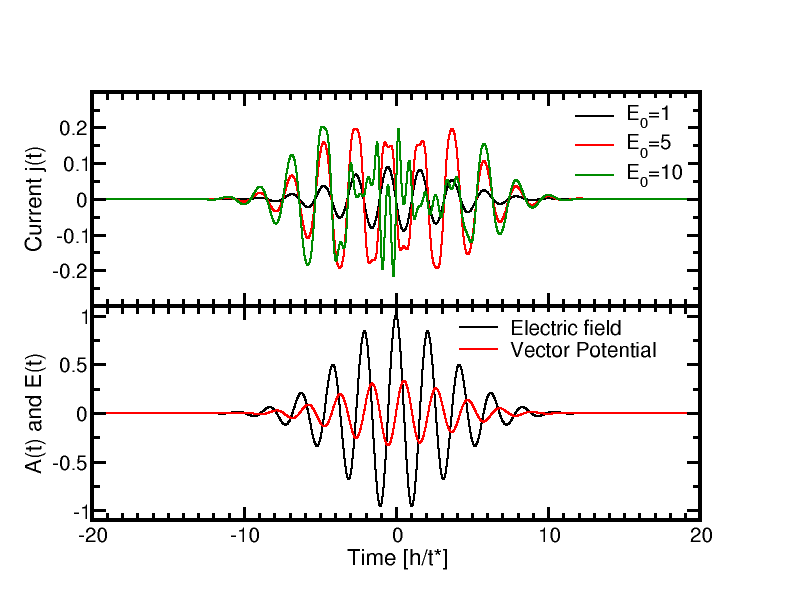}
\end{center}
\caption{(Color online.) Electrical current (top)  in the charge-density-wave insulator with $U=1$ and generated by the electric field pulse field pulse $\mathcal{E}(t)={E}_0\cos(3t)\exp (-t^2/25)$ (bottom)}
\label{fig: hhg_cosine}
\end{figure}

In our model, we study the Bloch electrons in a charge-density-wave material, which allows us to model both the excitation of the electrons into the upper band (and the creation of holes in the lower band) and the subsequent acceleration of those electrons and holes. As discussed before, the Peierls' substitution includes the electric field effects to all orders. Also since we don't have scattering in this system, but have dephasing, we are able to study high harmonic generation in this  system exactly.  Electromagnetic waves are produced by the accelerating charge, which is described by the time derivative of the electric current. So the light spectrum is proportional to 
\begin{equation}
\left | \int dt e^{i\omega t}\frac{d}{dt}\textbf{j}(t) \right |^2=|\omega J(\omega)|^2.
\end{equation}
Our method can calculate the transient nonequilibrium current under a strong electric field pulse in a charge-density-wave system. Note that we are assuming that the acceleration of electrons in the bulk dominates the production of light and surface effects are minimal.

\begin{figure}[!ht]
\begin{center}
\includegraphics[width=3.5in]{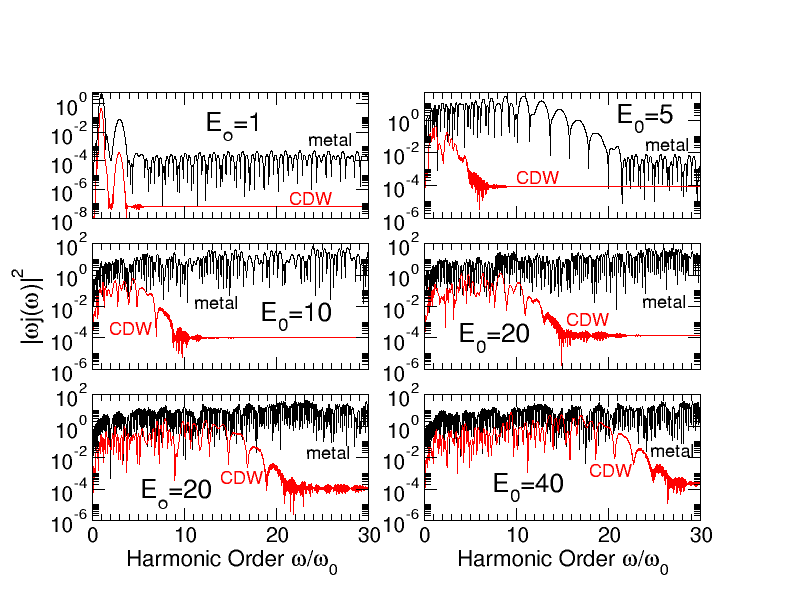}
\end{center}
\caption{(Color online.) High harmonic generation for the pulse $\mathcal{E}(t)={E}_0\sin(2t)\exp (-t^2/25)$ with $U=1$.
The red curve is for the charge density wave, while the black curve is for a metal \cite{JimHHG}. Note how the basic structure of the two responses are similar, but the metal has a much higher cutoff to the high harmonic spectra for the same amplitude. The overall amplitude is smaller for the charge density wave case, as expected, and the peak widths are somewhat narrower.}
\end{figure}
\begin{figure}[!ht]
\begin{center}
\includegraphics[width=3.5in]{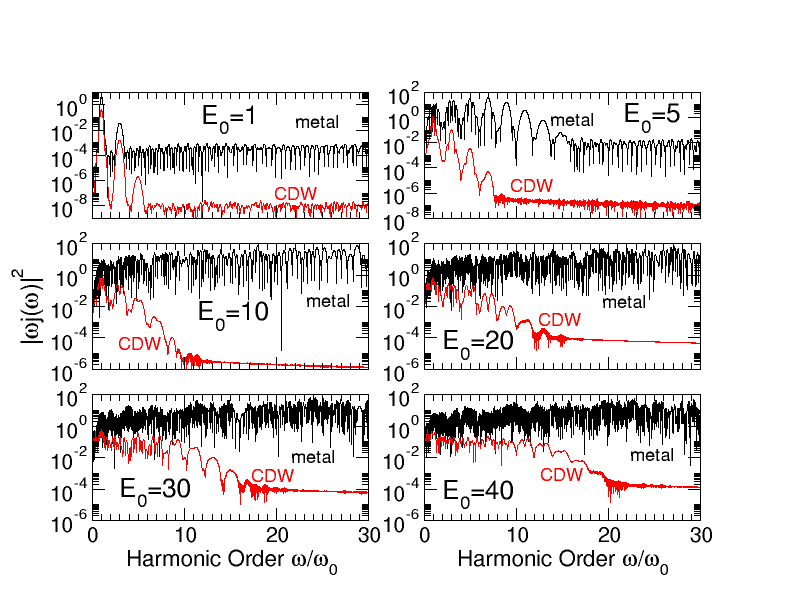}
\end{center}
\caption{(Color online.) High harmonic generation for the pulse $\mathcal{E}(t)={E}_0\cos(3t)\exp (-t^2/25)$ with $U=1$.
The red curve is for the charge density wave, while the black curve is for a metal \cite{JimHHG}.The behavior is similar to the lower frequency case, but with the difference in the cutoff frequency for the metal versus the insulator being even more striking. }
\end{figure}

Because the light pulse that is used in experiment is a propagating light pulse, it cannot have any dc component to it.  A typical pulse we use here is
\begin{equation}
\mathcal{E}(t)=E_0\sin(\omega_0t)\exp (-t^2/\sigma^2).
\end{equation}
Here $\sigma$ is a time constant that determines the pulse width. We can see that if $\sigma>2\pi /\omega_0$, the electric field pulse has a well defined frequency. 
Another pulse we will use in this work is
\begin{equation}
\mathcal{E}(t)=E_0\cos(3t)\exp (-t^2/\sigma^2).
\end{equation}
This pulse also yields nearly a zero vector potential for long times. So both pulses mimic the behavior of actual experimentally realizable pulses.

Figures \ref{fig: hhg_sin} and \ref{fig: hhg_cosine} show the electric field pulses we employ and the typical currents generated by them. This current is calculated for the case $U=1$ and $\sigma=5$.
For a small  amplitude ($E_0=1$), the current (the black line in the top panels) basically follows the vector potential, which is the time integral of the electric field. The sine pulse in Figure ~\ref{fig: hhg_sin} with $\omega_0=2$ corresponds to a 7-cycle pulse  and the cosine pulse in Figure \ref{fig: hhg_cosine} corresponds to a 9-cycle pulse. With such a low pulse amplitude,  the electron accelerates without hitting the Brillouin zone boundary, and hence, it corresponds to a nearly linear response regime. But when the field amplitude becomes larger, the current shows nonlinear behavior. In addition to the time-traces of the current, we also calculate the Fourier transform of the current to determine the radiated light spectrum.  This Fourier transform is performed over the time interval $[-40,80]$. 
Here due to the symmetry of a lattice (even under spatial inversions), the high harmonic spectrum produces only odd harmonics of the fundamental driving frequency \cite{JimHHG}. When the field amplitude becomes larger, we see more harmonics appear and the peaks for lower harmonics develop additional structure. It appears that the 9-cycle pulse case produces more recognizable peaks than the 7-cycle pulse, as expected. But for $E_0=1$, the 7-cycle pulse case produces a larger maximum HHG order.

\begin{figure}[!ht]
\begin{center}
\includegraphics[width=3.5in]{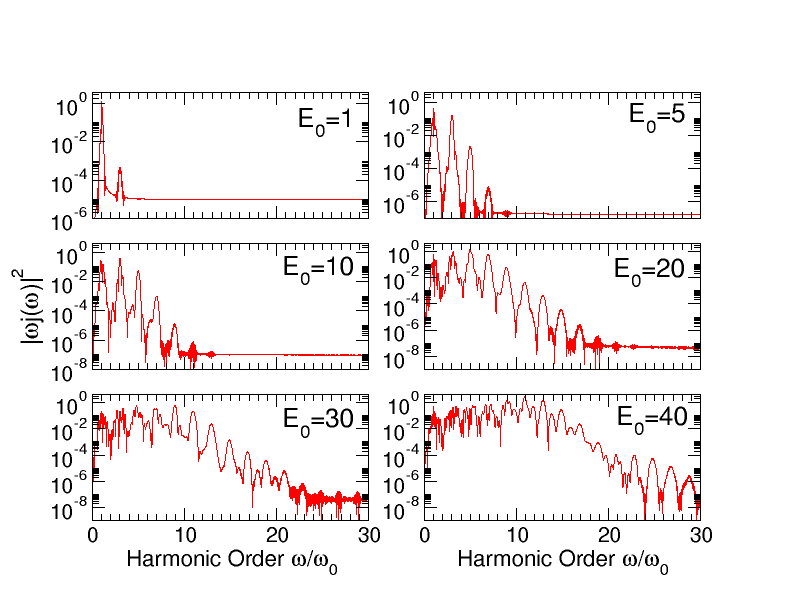}
\end{center}
\caption{High harmonic generation for the charge-density-wave insulator for the pulse $\mathcal{E}(t)=E_0\cos(3t)\exp (-t^2/25)$ and $U=3$.}
\label{fig: hhg_u=3}
\end{figure}

As we discussed above for ac driving fields, having $U$ equal to the electric field driving frequency $\omega_0$ enhances the generation of higher order harmonics. The corresponding high harmonic generation for $U=3$ with the 9-cycle pulse is shown in Figure \ref{fig: hhg_u=3}.
It is quite complicated how the charge-density-wave gap $U$ affects the high harmonic generation, as can be seen from a comparison between the $U=1$ and $U=3$ cases. By selecting $U$ equal to the driving frequency of the pulse, a larger maximum high harmonic generation order emerges, when compared to the $U=1$ case. The charge-density-wave gap $U$ also affects the additional structure in the lower harmonic peaks. For example, at $E_0=20$, a clear splitting is seen in the center  of each peak (for the low orders) in the $U=1$ case, but for the $U=3$ case, the peaks have not yet split.

One of the main differences with respect to the ac fields is that in the case of a pulse, we no longer see an additional modulation of the harmonic amplitudes as we saw in the ac driving case.  This is probably due to the smaller time range of the pulsed fields leading to broadened spectral features that prohibit destructive interference from occurring.  Nevertheless, we do see that for some field amplitudes, higher harmonics have a higher amplitude than the lower harmonics have.  This behavior arises primarily from the fact that the higher harmonics are generated when the field amplitudes are the largest, and hence this effect has a tendency to emphasize the higher harmonics.  Note further, that even though one cannot describe the photoexcitation process as occurring instantly at the point where the field amplitude is maximal, we do find that the results for the high harmonic generation spectra look remarkably similar to the results where one does not take into account the photoexcitation process \cite{JimHHG,lex} (except for more narrowing of the harmonic peaks in the charge density wave case and a smaller cutoff in the high harmonic generation for the same pulse amplitude). Hence, it appears that details of the photoexcitation process do not play a significant role in the generation of high harmonics.  The single most important characteristic is the amplitude of the pulsed field.
The gap $U$ introduces small differences in the additional structure peak splitting and an enhancement of the high harmonic generation when the driving frequency is resonant with the gap size ($\omega_0=U$).

\section{Conclusions}

In this work, we have developed an exact solution for the nonequilibrium problem of a charge-density-wave insulator placed in a large ac or pulsed electric field.  The theory is completely self-contained including all effects of the photoexcitation of electron-hole pairs and of their subsequent acceleration in the large electric field, incorporating the nonlinear effects of the electric field within an exact formalism. In the case of driving by an ac field, we saw that the most interesting nonlinear effects occur when the electric field amplitude is large, and that a Fourier transform of the temporal current traces display harmonics of the three important energy scales in the system, the driving frequency of the field, the gap of the charge density wave, and the energy gain due to the acceleration in the electric field over one lattice spacing. In addition, the regular, monochromatic nature of the ac driving field led to interesting beat-like structure for the strength of different harmonic peaks due to the possibility of destructive interference.  We also examined the effects of the field on the density of states for the ac field.

In the pulsed-field case, our main focus was on high harmonic generation.  Here, we found very similar behavior to that seen in single-band models that ignored the photoexcitation process. This includes how the strength of the peaks is fairly flat as a function of frequency until one hits a ``critical frequency'' where the peaks rapidly drop in strength.  The size of the critical frequency increases with the magnitude of the pulse amplitude and is larger in the metal than in the insulator.  In addition, as the amplitude is increased, we start to see complex structure develop within a single harmonic peak, starting for the lower harmonics. We also see circumstances where the peak magnitudes do not decay monotonically with increasing frequency, but there is a peak at some intermediate harmonic.  Because all of these features are seen in the single-band solutions, these results are strongly suggestive of the response having a universal character to it. The only significant difference that we see is that the peak widths tend to be more narrowed and the maximal high harmonics are reduced.

An important question, that we are not able to answer in this work, is how to optimize a material for use as a pulsed light source.  The main result we find with regards to optimization is that we want to have as small a Brillouin zone as possible, so that we can reach the nonlinear excitation regime with the smallest possible field amplitude.  This suggests that crystals with complex unit cells that include many atoms, may be better choices for high harmonic generation.  But the gap in the insulating phase suppresses the high harmonic generation as well, so a balance needs to be struck in a real material. Nevertheless, we hope that continuing experimental efforts will be able to see the universal features found in these calculations and will also be able to be controlled well enough that one can begin to develop devices that will utilize the high harmonic generation from solids in applications.

\acknowledgments
WS and JKF were supported for the development of the algorithm and coding by the National Science Foundation under grant number OCI-0904597.
JKF was supported by the US DOE, BES, MSED under grant number DEFG02-08ER46542 for the application of the algorithms to the high harmonic generation problem and by the McDevitt bequest at Georgetown University. 
AFK and TPD were supported by the US DOE, BES, MSED under contract number DEAC02-76SF00515. The collaboration
was supported by the US DOE, BES through the CMCSN program under grant number DESC0007091.
This work was made possible by the resources of the National Energy Research
Scientific Computing Center (via an Innovative and Novel Computational Impact on Theory
and Experiment grant), which is supported by the US DOE, Office of Science, under contract
number DE-AC02-05CH11231.

\end{document}